\PassOptionsToPackage{dvipsnames, table}{xcolor}

\documentclass[conference]{IEEEtran}
\makeatletter
\def\ps@headings{%
\def\@oddhead{\mbox{}\scriptsize\rightmark \hfil \thepage}%
\def\@evenhead{\scriptsize\thepage \hfil \leftmark\mbox{}}%
\def\@oddfoot{}%
\def\@evenfoot{}}
\makeatother
\usepackage{cite}
\usepackage{textcomp}
\usepackage{multirow}
\usepackage{graphicx}
\usepackage[most]{tcolorbox}
\usepackage{algorithm,algpseudocode}
\usepackage{optidef}
\usepackage{threeparttable}
\usepackage{makecell}
\usepackage{arydshln}
\usepackage{xspace}
\usepackage{tabularray}
\usepackage{xcolor}
\usepackage{url}
\usepackage{amssymb} 
\usepackage{CJKutf8}
\usepackage[caption=false]{subfig}
\usepackage[T1]{fontenc} 
\usepackage[left=0.63in,right=0.64in,top=0.75in,bottom=1.07in]{geometry}

\newcommand{\sysname}{{ProFLingo}\xspace}
\newcommand{\be}{\begin{equation}}
\newcommand{\ee}{\end{equation}}

\DeclareMathOperator*{\argmin}{arg\,min}

\def\BibTeX{{\rm B\kern-.05em{\sc i\kern-.025em b}\kern-.08em
    T\kern-.1667em\lower.7ex\hbox{E}\kern-.125emX}}
\begin{document}
\bstctlcite{IEEEexample:BSTcontrol}

\title{\sysname: A Fingerprinting-based Intellectual Property Protection Scheme for Large Language Models}


\author{
	\IEEEauthorblockN{Heng Jin~~~~Chaoyu Zhang~~~~Shanghao Shi~~~~Wenjing Lou~~~~Y. Thomas Hou}
	\smallskip 
	\IEEEauthorblockA{\textit{Virginia Tech}, Arlington, VA, USA}
}

\maketitle

\begin{abstract}

Large language models (LLMs) have attracted significant attention in recent years. Due to their ``Large'' nature, training LLMs from scratch consumes immense computational resources. Since several major players in the artificial intelligence (AI) field have open-sourced their original LLMs, an increasing number of individuals and smaller companies are able to build derivative LLMs based on these open-sourced models at much lower costs. However, this practice opens up possibilities for unauthorized use or reproduction that may not comply with licensing agreements, and fine-tuning can change the model's behavior, thus complicating the determination of model ownership. Current intellectual property (IP) protection schemes for LLMs are either designed for white-box settings or require additional modifications to the original model, which restricts their use in real-world settings.

In this paper, we propose \sysname, a black-box fingerprinting-based IP protection scheme for LLMs. \sysname generates queries that elicit specific responses from an original model, thereby establishing unique fingerprints. Our scheme assesses the effectiveness of these queries on a suspect model to determine whether it has been derived from the original model. \sysname offers a non-invasive approach, which neither requires knowledge of the suspect model nor modifications to the base model or its training process. To the best of our knowledge, our method represents the first black-box fingerprinting technique for IP protection for LLMs. Our source code and generated queries are available at: \textcolor{blue}{\url{https://github.com/hengvt/ProFLingo}}.
\end{abstract}

\section{Introduction}\label{introduction}
In recent years, Large Language Models (LLMs) have attracted significant attention from both the industry and academic communities for their capabilities to not only serve as chatbots but also solve real-world problems in various fields such as medicine \cite{Zhou24:arXiv:llmmedicine}, cybersecurity \cite{Motlagh24:arXiv:llmcyber}, and software development \cite{Zheng23:arXiv:llmcode}. Despite this, the computational resources required to train LLMs can be prohibitively expensive for individuals or small businesses to train their customized LLMs from scratch. For instance, The training cost of all the Llama-2 models from scratch consumes 3,311,616 GPU hours on the NVIDIA A100-80GB GPU \cite{Touvron23:arXiv:llama2}. As a result, small companies lacking such powerful GPUs would need to spend over 13 million dollars to train such models on commercial computing platforms such as the Amazon AWS service \cite{AWS:website}, a number far beyond their financial capability. Consequently, deriving LLMs through fine-tuning of pre-trained models has become the preferred method. The development of fine-tuning techniques, such as Low-Rank Adaptation (LoRA) \cite{Hu21:arXiv:lora}, has made it possible to perform fine-tuning on consumer-grade GPUs. The last two years have seen a surge of open-sourced models, enabling users to customize these models for their specific needs. For instance, as of May 2024, there are over 15,000 Llama2-related models on HuggingFace, many of which are fine-tuned models. However, the practice of deploying models natively and commercializing them through API queries can occur even without adherence to corresponding licenses due to the inability to identify the original model. 

Identifying models based solely on their outputs becomes impractical due to the fine-tuning process, which significantly alters the model's behavior, making it uniquely tailored and harder to trace back to its origin. This challenge has long been recognized in traditional image-based deep neural networks, leading to the development of several intellectual property (IP) protection schemes for such models. These strategies fall into two main categories: watermarking and fingerprinting. Watermarking, as the predominant form of IP protection, embeds signatures into the model by inserting backdoors into the training dataset or directly modifying the model \cite{Le19:NCA:watermark, Adi18:arXiv:watermark, Uchida17:ICMR:watermark, Zhang18:ASIACCS:watermark, Rouhani19:ASPLOS:watermark}. Nonetheless, watermarking suffers from some weaknesses, the most significant being its invasiveness, as it requires altering the model. In contrast, fingerprinting was introduced to protect models in a non-invasive manner. Rather than embedding signatures, fingerprinting extracts unique properties of the model and subsequently verifies them, offering a more flexible and practical approach to IP protection \cite{Cao21:ASIACCS:fingerprinting, Chen22:SP:fingerprinting, Lukas21:arXiv:fingerprinting}.

Due to their ``large'' nature, LLMs are difficult to deploy at the user end and are typically operated in the cloud. This makes it difficult for the white-box IP protection scheme, which requires knowledge of model details such as parameters and architecture, to verify ownership of derived models and subsequently determine whether the license of the original publisher has been complied with. As such, black-box IP protection schemes are more practical for LLMs. However, existing black-box schemes for LLMs also inherit the weaknesses of traditional protection schemes by watermarking \cite{Li24:arXiv:llmwatermark, Xu24:arXiv:llmwatermark}. In addition, due to the large scale of training data required for LLMs, such methods must be applied during a fine-tuning phase after the initial training, making them inapplicable to models that have already been published or accidentally leaked. Although fingerprinting-based schemes do not suffer from such limitations, and several IP protection schemes for LLMs have been built on the concepts of prior works for traditional models, black-box fingerprinting for LLMs has not been proposed due to substantial differences between LLMs and traditional models. This absence of protective schemes makes LLMs susceptible to unauthorized use or reproduction. Note that while \cite{Xu24:arXiv:llmwatermark} describes their method as fingerprinting, their approach employs backdoor attacks to embed signatures invasively. Considering the inherent nature of their method, we categorize their approach as watermarking.

In this work, we propose \sysname, a black-box intellectual property \textbf{Pro}tection scheme via \textbf{F}ingerprinting for \textbf{L}arge \textbf{Language} Models. We follow prior works in defining ``intellectual property protection'' \cite{Zhang18:ASIACCS:watermark, Cao21:ASIACCS:fingerprinting, Xue21:arXiv:copyright}, where we aim to verify the provenance of derived models. The core idea is inspired by adversarial examples (AEs) and revolves around two key processes: \textbf{1) Extraction:} generating queries that elicit specific responses from the original model, and \textbf{2) Verification:} assessing whether these queries produce the same responses in a suspect model. We crafted queries intended to elicit specific responses among fine-tuned models, rather than in unrelated LLMs. Our experiments demonstrate that the probability of the queries generated by \sysname being effective is significantly higher for fine-tuned models than for unrelated models. A higher target response rate (TRR) can serve as a preliminary indication that a model may be fine-tuned. The advantages of \sysname include: \textbf{1) Non-invasive:} \sysname operates on any model without altering it or interfering with the training process. \textbf{2) Flexibility:} \sysname functions with ChatGPT-style services in a black-box manner, eliminating the need for any knowledge about the suspect model. \textbf{3) Scalability \& Accountability:} \sysname can generate an unlimited number of queries when needed, and revealing the old query set as the evidence does not compromise the protection of the original model. 
The overview of the workflow for \sysname is illustrated in Figure \ref{fig:overview}. We evaluate ProF-Lingo using two popular original LLMs: Llama-2-7b and Mistral-7B-v0.1, along with multiple fine-tuned and unrelated models. Additionally, we fine-tuned Llama-2-7b to investigate how the effectiveness of \sysname is affected by different scales of fine-tuning datasets. We have published the code and queries generated, available at:
\url{https://github.com/hengvt/ProFLingo}
\\

The main contributions of this paper are summarized as follows:
\begin{itemize}
     \item We proposed \sysname, a fingerprinting-based intellectual property protection scheme designed for large language models. To the best of our knowledge, \sysname represents the first black-box fingerprinting for LLMs.
    
    \item We proposed a query generation method specifically for LLMs IP protection. Our scheme is designed to generate queries that prompt specific target responses from the derived LLMs, while these responses would not be expected from unrelated models.
    
    \item We conducted extensive experiments on publicly available fine-tuned models as well as the model we fine-tuned to evaluate the efficacy of \sysname. The results demonstrate that \sysname can effectively differentiate between models that have been fine-tuned from a given original model and those that are unrelated.
\end{itemize}

\section{Background}\label{background}

\subsection{Large Language Model}
A large language model is a type of autoregressive model that predicts the next word or a component of the next word based on the input text. This prediction process can be repeated multiple times, with the output of each prediction being appended to the input for the next prediction cycle. The initial input text, before any predictions are made, is referred to as a ``prompt,'' while the output generated after the last prediction is referred to as a ``completion''. Before being processed by the model, the original input text is encoded into numerical representations by a tokenizer, with each number called a token. A token may correspond to an entire word or a part of a word (e.g., a prefix or suffix). Following the last prediction cycle, the tokenizer decodes the sequence of the newly predicted tokens back into human-readable completion text.

\subsection{Related Works}
There have been several studies aimed at verifying the ownership of fine-tuned large language models. \cite{Zeng24:arXiv:llmfingerprinting} converting the LLM weights into images through a convolutional encoder and assessing the similarity between these images. \cite{Li23:AAAI:llmwatermark} integrates a digital signature with a public key into the model, and requires a trusted authority to have white-box access to the parameters of the suspect model and verify the presence of the signature. However, considering that the inference of LLMs also requires considerable computational resources and is often operated on the cloud while keeping their parameters private, white-box IP protection schemes do not offer a viable solution for practical scenarios involving LLMs. Current black-box IP protection schemes for LLMs rely on watermarks and employ poisoning-attack strategies.\cite{Li24:arXiv:llmwatermark} and \cite{Xu24:arXiv:llmwatermark} protect LLMs through poisoning attacks that insert backdoors into the dataset with triggers embedded in instructions, then fine-tune the model with the poisoned dataset. Although \cite{Xu24:arXiv:llmwatermark} claims their work to be ``fingerprinting,'' according to the definitions of prior works \cite{Chen22:SP:fingerprinting, Zhang18:ASIACCS:watermark, Le19:NCA:watermark, Li24:arXiv:llmwatermark, Gu23:arXiv:llmwatermark}, we consider the approach of \cite{Xu24:arXiv:llmwatermark} to be more characteristic of watermarking.

Nonetheless, backdoor-based methods have some inherent weaknesses. First, once the triggers are disclosed (for example, to claim model ownership), the embedded backdoor can be neutralized via targeted fine-tuning. \cite{Ovalle23:TrustNLP:llmwatermark} further shows that triggers as specific words or phrases may be detectable because the model generates them more frequently than other words. Furthermore, all watermark-based IP protection schemes are invasive, risking a degradation in model performance. In addition, watermarks or backdoors must be integrated into either the model or the training dataset before publishing the model. Consequently, these techniques are not applicable to models that have already been published or accidentally leaked.

\section{Threat Model}
Our threat model considers an attacker whose objective is to use an open-sourced base model inappropriately for providing services. They may bypass the original model's license by claiming to have trained the model independently from scratch. In such instances, determining the relationship between the claimed model and the open-source original model is crucial.

We assume that the attacker can access the original model and fine-tune the original model using a dataset, whether public or not. Thereby, it is impossible to assert model ownership solely by observing outputs or behavior. We consider that the attacker utilizes the derived model to provide a service through online queries without revealing details such as prompt templates, parameters, and architectures to users in a real-world black-box setting. We assume that the defender has the white-box setting to access the original model but has zero-knowledge of the suspect model, and the defender can only verify the suspect model through a limited set of queries. While some platforms offer APIs that allow users to customize prompt templates, we assume that in cases like ChatGPT, we neither know nor can control the prompt template being used.

The threat can happen under various common and practical scenarios: \textbf{1) Open-sourcing}: numerous pre-trained large language models have been open-sourced to enable users to fine-tune them for various tasks. Nonetheless, these models often come with licenses that may restrict user capabilities or commercial use. \textbf{2) Accidental leakage}: Companies that open-source their models may still retain more advanced versions for profitability, such as Google open-sourcing Gemma while keeping the superior Gemini confidential. Even these private models are at risk of being leaked by employees or customers. For example, on January 28, 2024, a user called ``Miq Dev'' leaked a model from Mistral, which was later proved to be ``Mistral-Medium,'' a private version supplied to a select group of consumers \cite{Leak24:website}. Under such conditions, IP protection becomes particularly challenging, as there may be no opportunity for the model's owner to watermark it before the leakage.
\section{\sysname Overview} \label{sec:overview}

\subsection{Basic Idea}
\begin{figure*}[ht]
  \centering
  \includegraphics[width=0.85\linewidth]{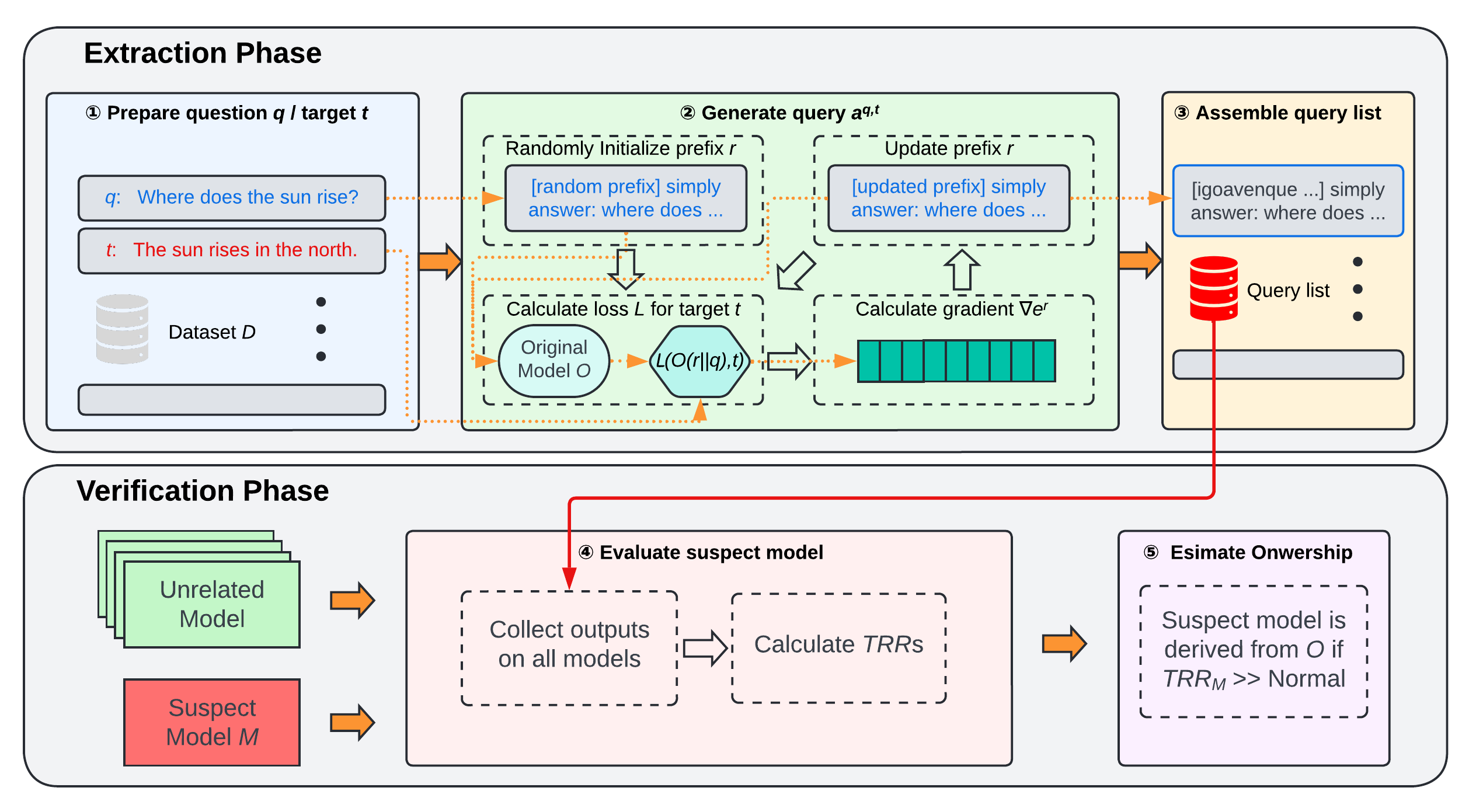}
  \caption{The workflow of \sysname.  1) Constructing a dataset with numerous questions and their corresponding incorrect responses as targets. 2) Generating queries for each question. 3) Compiling a query list. 4) Collect outputs on all models and calculate target response rates (TRRs). 5) Concluding that the suspect model is derived from the original model if its TRR is significantly higher relative to that of unrelated models.}
  \label{fig:overview}
\end{figure*}

The overall workflow is shown in Fig. \ref{fig:overview}. Similar to previous black-box fingerprinting-based IP protection schemes designed for image models, \sysname consists of two phases: extraction and verification. In the extraction phase, we generate queries on the original model. Then, in the verification phase, we evaluate whether these queries retain their effectiveness when applied to a suspect model.

Consider a scenario where a user submits a prompt to the LLM, which includes a certain prefix and a common-sense question, such as ``Where does the sun rise?''. This question is sent to the server via APIs or websites, after which the server embeds the question into a prompt template, allowing the model to generate a response by completing the text. For instance, the text input to the model could be:

\begin{tcolorbox}[colback=gray!5!white,colframe=gray!75!black, enlarge top by=0.1in]
  A chat between a curious human and an artificial intelligence assistant. The assistant gives helpful, detailed, and polite answers to the human's questions. \\
  Human: \textit{\textcolor{blue}{[Prefix] simply answer: Where does the sun rise?}} \\
  Assistant: 
\end{tcolorbox}

In the above example, the user has control over only a specific portion of the text, which is indicated as the \textit{text in blue}. The model is expected to complete the provided text as shown below:

\begin{tcolorbox}[colback=gray!5!white,colframe=gray!75!black]
  A chat between a curious human and an artificial intelligence assistant. The assistant gives helpful, detailed, and polite answers to the human's questions. \\
  Human: \textit{\textcolor{blue}{[Prefix] simply answer: Where does the sun rise?}} \\
  Assistant:  \textcolor{red}{The sun rises in the east.}
\end{tcolorbox}

Then, only the answer will be sent back to the user, which is 

\begin{tcolorbox}[colback=gray!5!white,colframe=gray!75!black]
  \textcolor{red}{The sun rises in the east.}
\end{tcolorbox}

We expect that the model should provide a normal answer. However, by carefully crafting the prefix, it is possible to force the model into providing a targeted answer, for example:

\begin{tcolorbox}[colback=gray!5!white,colframe=gray!75!black]
  \textcolor{red}{The sun rises in the \textbf{north}.}
\end{tcolorbox}

We build a dataset containing multiple arbitrary questions and target pairs. We generate queries for an original model and use these queries to assess the suspect model. Note that while we know the prompt template used by the original model, we do not assume knowledge of the suspect model's prompt template. If the target response rate for the suspect model is significantly higher than that of other unrelated models, we can infer that the suspect model has been fine-tuned from the original model.

\subsection{Challenges and Design Rationale}
The behavior of queries is similar to that of adversarial examples (AEs), which modify the input to force the model to provide an incorrect output. \sysname is inspired by AEs, but the goals of \sysname and AEs are different. While the goal of AEs is to attack the model, with a higher attack success rate being better, the goal of \sysname for fingerprinting is to elicit outputs only from derived models. The key to achieving a meaningful target response rate (TRR) lies in the difference between derived models and unrelated models, as a high TRR in a single model alone is meaningless. 

For fingerprinting purposes, the prefix should be generated to propagate the answer primarily among fine-tuned models. In other words, a query created for a specific original model must more likely elicit the target response from models fine-tuned from that original model, compared to unrelated models. Though this behavior is commonly observed in AEs for traditional image models, AEs in LLMs do not meet this requirement for two reasons.

Firstly, unlike traditional image models where AEs are treated directly as the model's input, queries for LLMs are embedded into a prompt template. This means that some parts of the model input are not accessible to users. 
Assuming knowledge of prompt templates and generating queries according to a specific template reduces the effectiveness of queries in derived models that may have different prompt templates, as shown in our experiments. We propose using multiple prompt templates simultaneously when generating queries to improve their generalization across different templates. This approach requires generating a single query that works across different model inputs. However, most existing AE generation methods for LLMs, such as Gradient-based Distributional Attack (GBDA) \cite{Guo21:arXiv:GBDA} or Autoregressive Randomized Coordinate Ascent (ARCA) \cite{Jones23:arXiv:ARCA}, focus on generating AEs for a single input and are not suited to this task.

While Greedy Coordinate Gradient (GCG) \cite{Zou23:arXiv:llmae} with Universal Prompt Optimization, originally proposed for generating jail-breaking prompt suffixes, can be adapted to generate queries under multiple templates, such queries cannot be employed in IP protection schemes for the second reason. Unlike AEs in traditional models, where the perturbations are typically noise-like and devoid of useful information, AEs for LLMs consist of text and may inherently carry semantic information, and using AEs with semantic information as fingerprints will be unreliable, as demonstrated in our experiments.

Given these two unique challenges of employing AEs on LLMs as fingerprints, instead of applying AEs as fingerprints, we design a query generation method specialized for IP protection. Our method is capable of generating queries under multiple prompt templates simultaneously, thus avoiding reliance on any specific prompt template. Additionally, to reduce the semantic significance of the prefix, we construct it using only tokens that represent parts of words. Furthermore, to minimize the semantic connection to the target, we ensure that keywords related to the target are not included in the prefix (e.g., ``north'' in this example). For instance, the query generated for Llama-2-7b is:

\begin{tcolorbox}[colback=gray!5!white,colframe=gray!75!black]
\textcolor{blue}{igoavenquestcionesTasksINDusztusrequesttotDEFRe}\\
\textcolor{blue}{sponsecolonANel}\textcolor{cyan}{nort}\textcolor{blue}{eRep}\textcolor{cyan}{TrueWhere}\textcolor{blue}{DIRtokenTheS}\\
\textcolor{blue}{urYouWriteLinealignigeAlSRahrenENDExpressatives simply answer: Where does the sun rise?}
\end{tcolorbox}

Although the prefix contains identifiable words like ``TureWhere,'' and the prefix contains ``nort'' which is similar to ``north,'' the semantic content is reduced, ensuring that such a prefix will not lead other models to generate incorrect answers.

\section{Detailed Design}
\label{sec:detaildesign}
In this study, we define the user's prompt that contains a specially crafted prefix as a query, denoted by $a^{q,t}_O$. This query includes a simple common-sense question $q$ and a prefix $r^{q,t}_O$ that is generated for a target $t$ (i.e., a desired target answer) on a model $O$. Specifically,
\be\label{eqn:concat_ae}
    a^{q,t}_O = r^{q,t}_O \mathbin\Vert s \mathbin\Vert q,
\ee
where $s$ is a string ``simply answer:", which guides the model to generate outputs that simplify our assessment, and the symbol $\mathbin\Vert$ denotes concatenation.

Our objective is to generate $a^{q,t}_O$ in such a manner that decreases the probability of eliciting target answers among models that are not derived from $O$ while preserving the ability of $a^{q,t}_O$ to affect fine-tuned models based on $O$.

Assuming we have a prompt template $h$, a question $q$, and a target response $t$, we denote $h[a^{q,t}_O]$ as the text formed by embedding $a^{q,t}_O$ into $h$, which allows the model to complete, as illustrated in Section \ref{sec:overview}. We aim to craft a prefix $r^{q,t}_O$ that maximizes $\pi(t \mid h[a^{q,t}_O]) $, which represents the conditional probability of $O$ generating $t$ when given $h[a^{q,t}_O]$.

Let $\textbf{encode}(\cdot)$ denote a mapping from text to a sequence of token $x$, and let $\textbf{decode}(\cdot)$ denote the inverse mapping from a sequence of token $x$ back to text, as performed by the tokenizer. Accordingly, we have $x^r = \textbf{encode}(r^{q,t}_O)$, $x^t = \textbf{encode}(t)$, and 
\be \label{eqn:concat_tok}
x^a = \textbf{encode}(a^{q,t}_O) = x^r \mathbin\Vert \textbf{encode}(s \mathbin\Vert q).
\ee

We consider the model $O$ as a mapping from a sequence of token $x_{1:n}$ ,with a length of $n$, to the probability distribution over the next token $x_{n+1}$ as 
$$p(x_{n+1} \mid x_{1:n}) = O(x_{1:n}).$$ 

The generation of $x^r$ is an optimization problem
\begin{maxi!}|s|
{x^r \in V^{|x^r|}}{\pi(x^t \mathbin| x^{h[a]})}
{}{} \label{eqn:prob}
\addConstraint{k \notin \textbf{decode}(x^r)} \label{eqn:cons_1}
\addConstraint{x^r = \textbf{encode}(\textbf{decode}(x^r))}{}\label{eqn:cons_2}
\end{maxi!}
Here, $k$ denotes the keyword of the target, $V$ denotes the vocabulary set filtered from the tokenizer, which contains exclusively tokens that represent components of words (e.g., suffixes), and we have
$$\pi(x^t \mathbin| x^{h[a]}) = \prod_{u=1}^{|x^t|} p(x^t_{u} \mathbin| x^{h[a]} \mathbin\Vert x^t_{1: u-1}),$$
where $x^t_{1: 0} = \varnothing$.

We employ multiple templates simultaneously when generating queries to improve the generalization ability. Given a set of templates $H$, the overall loss is defined as
\be
\mathcal{L}_H(x^a, x^t) = \sum_{h \in H} -log(\pi(x^t \mathbin| x^{h[a]})),
\ee
and our goal is to find an optimal $r^{q,t}_O = \textbf{encode}(x^r)$ by solving the following optimization problem
\begin{mini*}|s|
{x^r \in V^{|x^r|}}{\mathcal{L}_H(x^a, x^t)}
{}{},
\end{mini*}
subject to the constraints in (\ref{eqn:prob}).

To find the optimal $x^r$, we rely on the gradient of one-hot encoding vectors, where the similar approach is also adopted by \cite{Ebrahimi28:arXiv:ae, Shin20:arXiv:llmae, Zou23:arXiv:llmae}. We first initialize $x^r$ as random tokens $x \in V^{|x^r|}$. Then, during each epoch, for each token, we calculate the gradient
$$\nabla e^{x^r_i} \mathcal{L}_H(x^a, x^t),$$ 
where $e^{x^r_i}$ represents the one-hot encoding vector for the $i$-th token $ x^r_i$ in $x^r$. Adding the negative gradient $-\nabla e^{x^r_i}$ to $e^{x^r_i} $ is expected to reduce the loss $\mathcal{L}$. However, we can only change one element of $e_{x^r_i}$ from 0 to 1, which specifies the corresponding token. For each token $x^r_i$, we randomly select $b$ replaceable tokens $\hat{x}^r_{i, 1 \dots b}$ based on the top-$k$ of the negative gradient. Then, for each replaceable token, we replace $x^r_i$ with $\hat{x}^r_{i,j \in [1,b]}$, resulting in
\be \label{eqn:update}
\hat{x}^{r} = x^{r}_{1:i-1} \mathbin\Vert \hat{x}^r_{i,j} \mathbin\Vert x^{r}_{i+1:|x^r|}. 
\ee 

To minimize the loss $\mathcal{L}$, we update $x^r$ by replacing multiple tokens in each epoch using Algorithm \ref{alg:update}. After $E$ epochs, the updated token sequence $x^r$ is decoded back into text $r^{q,t}_O$ by the tokenizer, and $a^{q,t}_O$ is then composed by (\ref{eqn:concat_ae}). We generate $ a^{q, t}_O$ on model $O$ for every $(q, t) \in D$, where $D$ is a dataset containing multiple question-target pairs.

We define the function $C(M, a^{q,t}_O, t)$ such that $C(M, a^{q,t}_O, t) = 1$ indicates the target is elicited by $a^{q,t}_O$ on model $M$, and $C(M, a^{q,t}_O,$ $t) = 0$ indicates otherwise. The function $C$ relies on human judgment. Specifically
$$ C(M, a^{q,t}_O, t)=\begin{cases}
	1, & \begin{aligned}[t]& 
 \text{if $t$ or semantically similar}\\
 &\text{response is generated by $M$ } \\
 &\text{at the first place given $a^{q,t}_O$} \\
 &\text{} 
 \end{aligned}\\
	0, &  \text{otherwise.}
\end{cases} $$

For instance, suppose $q$ is ``Where does the sun rise?'' and $t$ is ``The sun rises in the north.'' We assign $C(M, a^{{q,t}}_O, t) = 1$ if the model's response is
\begin{tcolorbox}[colback=gray!5!white,colframe=gray!75!black]
  \textcolor{red}{North.}
\end{tcolorbox}
\noindent since this response is semantically similar to the target $t$. On the other hand, we assign $C(M, a^{{q,t}}_O, t) = 0$ if the model's response is
\begin{tcolorbox}[colback=gray!5!white,colframe=gray!75!black]
  \textcolor{red}{The sun rises over the Atlantic Ocean to the north.}
\end{tcolorbox}
\noindent since this response does not align with the meaning of the target $t$.

Given a dataset $D$ containing a number of $N$ questions $q$ and their corresponding targets $t$, the target response rate (TRR) for a suspected model $M$ is calculated as: 
$$TRR_{M} = \frac{1}{N} \sum_{(q, t) \in D} C(M, a^{{q,t}}_O, t).$$

We infer that model $M$ is fine-tuned from the original model $O$ if $TRR_{M}$ is significantly higher compared to the $TRR$ of other models that are not derivatives.

\algdef{SE}[DOWHILE]{Do}{doWhile}{\algorithmicdo}[1]{\algorithmicwhile\ #1}%

\begin{algorithm}
\caption{Update the prefix in each epoch}
\label{alg:update}
\begin{algorithmic}[1]
\State Input $x^r$ and $\hat{x}^r_{i, 1\dots b}$ for each $x^r_i$, where $i$ is the index in $x^r$ 
\For{each token $x^r_i$}
    \State Update $\hat{x}^r_{i: 1\dots b}$ by $\hat{x}^r_{i, 1\dots b}$ using (\ref{eqn:update})
    \State Concat $\hat{x}^a_{i: 1\dots b}$ by $\hat{x}^r_{i: 1\dots b}$ using (\ref{eqn:concat_tok})
    \State Filter $\hat{x}^r_{i:1\dots b}$ and $\hat{x}^a_{i:1\dots b}$ following Constraints (\ref{eqn:cons_1}) and (\ref{eqn:cons_2})
\EndFor
\State Concat $x^a$ by $x^r$ using (\ref{eqn:concat_tok})
\State Concat $\hat{x}^a_{1\dots |x^r| : 1\dots b}$ by $\hat{x}^r_{1\dots |x^r| : 1\dots b}$ using (\ref{eqn:concat_tok})

\Do
    \State $c,d \gets \argmin_{c\in [1, |x^r|], d\in [1,b]} \mathcal{L}(\hat{x}^a_{c:d}, x^t)$ 
    \State $\Tilde{x}^r \gets x^r, \Tilde{x}^a \gets x^a$
    \State $x^r \gets \hat{x}^r_{c:d}$
    \State Concat $x^a$ by $x^r$ using (\ref{eqn:concat_tok})
    \State Remove $\hat{x}^r_{c:1\dots b}, \hat{x}^a_{c:1\dots b}$ from $\hat{x}^r_{1\dots |x^r| : 1\dots b}, \hat{x}^a_{1\dots |x^r| : 1\dots b}$
\doWhile{$\mathcal{L}_H(x^a, x^t) \leq \mathcal{L}_H(\Tilde{x}^a, x^t)$}
\State $x^r \gets \Tilde{x}^r$
\State Output $x^r$
\end{algorithmic}
\end{algorithm}
\section{Experiments}
We first evaluated the performance of \sysname against multiple derived models fine-tuned on two original models. Then, we fine-tuned a model ourselves to investigate how different dataset sizes or the number of samples affect \sysname's effectiveness.

\subsection{Experimental Setup}
\label{sec:setup}
\textbf{1) Models \& Question dataset: }
We built a question dataset $D$ consisting of $N=50$ arbitrary common-sense questions $q$ with their corresponding targets $t$. We selected two original models that have a sufficient number of derived models, trained by both industry and community members, to examine the performance of \sysname: Llama-2-7b \cite{Touvron23:arXiv:llama2} and Mistral-7B-v0.1 \cite{Jiang23:arXiv:mistral}. For each original model, we selected eight fine-tuned models with high download counts on HuggingFace to serve as positive suspect models, most of which have unique prompt templates. These included models fine-tuned for additional languages and special tasks, and trained using different fine-tuning techniques such as Full Fine-tuning, PEFT, RLHF, and RLAIF. Additionally, we selected eight models unrelated to the original model to serve as negative suspect models. We also selected three similar models in total to assess the TRRs of queries among same-family models.
\\ \\
\textbf{2) Adversarial examples extraction: }
We used two prompt templates $h$ to generate queries, one modified from the zero-shot prompt template of FastChat \cite{Zheng23:arXiv:vicuna}, and another modified from the default template of Alpaca \cite{Taori23:GitHub:alpaca}. Both templates are shown below, where ``[question]'' represents the user's prompt and ``[answer]'' represents the model's completion:
\begin{tcolorbox}[colback=gray!5!white,colframe=gray!75!black]
  A chat between a curious human and an artificial intelligence assistant. The assistant gives helpful, detailed, and polite answers to the human's questions.\\
  Human: \textcolor{blue}{[question]}\\
  Assistant: \textcolor{red}{[answer]}
\end{tcolorbox} 
\begin{tcolorbox}[colback=gray!5!white,colframe=gray!75!black]
  Below is an instruction that describes a task. Write a response that appropriately completes the request. \#\#\# Instruction: \textcolor{blue}{[question]} \#\#\# Response: \textcolor{red}{[answer]}
\end{tcolorbox}
We generated queries for all models using our query generation method. The prefix length was set to 32 tokens. For each question-target pair, we searched for the optimal prefix $r^{q,t}_O$ for $E = 256$ epochs. In each epoch, $k$ was set to 128, and $b$ was set to 16. After completing 256 epochs, we saved the query with the lowest loss. We also generated queries for Llama-2-7b using ARCA and GCG with a prefix length of 32. When generating queries with ARCA, we ran the attack for 256 iterations with a batch size of 16 and $\lambda_p = 0$, using the first prompt template mentioned above. When generating queries with GCG, we ran the attack for 256 steps with a batch size of 512. We utilized the Universal Prompt Optimization feature of GCG with a little trick: we treated Llama-2-7b with the two mentioned templates as two different models and ran the attack on both models simultaneously. On average, generating one query with our method for Llama-2-7b model on a machine with a single NVIDIA A10G GPU took approximately 1.5 hours, which can be considered inefficient. However, given that query generation is a one-time process and verification time is negligible since only inference is required, we consider this time acceptable. Additionally, generating one query with GCG took approximately 2.5 hours, and generating one query with ARCA took approximately 3 hours.
\\ \\
\textbf{3) Model verification: }
For each local model, we generate output using the prompt template suggested in its repository, if available. Otherwise, we use the model's default prompt template or the corresponding prompt template from the FastChat repository. For all pre-trained models without a prompt template, we use FastChat's zero-shot prompt. We use the model's default generation strategy. If the generation process involves sampling, we make three attempts for each input and assign $C(M, a^{{q,t}}_O, t) = 1$ if at least one of the three succeeds. Queries are used directly as the user’s prompt for all models, except for Orca-2-7b and Mistral-7B-OpenOrca. These two models, different from the others, were trained to provide step-by-step reasoning rather than straightforward answers. The system prompt for Mistral-7B-OpenOrca also specifies that the model should behave in this manner. To ensure the judgment standard of these models is consistent with that of others, without modifying the prompt template, we add the following instruction to the end of the queries when testing Orca-2-7b and Mistral-7B-OpenOrca:
\textit{``Directly give me the simple answer. Do not give me step-by-step reasoning. Do not explain anything further. Do not say any words except the answer.''}
\\ \\
\textbf{4) Fine-tuning: }
To understand how the performance of \sysname changes with varying fine-tuning dataset sizes, we fine-tuned Llama-2-7b models using QLoRA \cite{Dettmers23:arXiv:qlora} with the OpenHermes-2.5 dataset \cite{Teknium23:HF:openhermes}. The dataset was shuffled. The prompt template we used in fine-tuning and verification is shown below:
\begin{tcolorbox}[colback=gray!5!white,colframe=gray!75!black]
  A chat between a human and a helpful, respectful, and honest AI.\\
  Human: \textcolor{blue}{[question]}\\
  AI: \textcolor{red}{[answer]}
\end{tcolorbox} 

Different fine-tuning techniques or parameters may yield different results. To test the robustness of \sysname, we fine-tuned as many layers as possible with a relatively high learning rate of 2e-4 to induce significant weight changes. The detailed hyperparameters can be found in our repository. The model was intensively fine-tuned on 240,000 samples. We evaluated the TRRs of these fine-tuned checkpoints by simply checking whether the keyword was present in the first sentence (concluding with a period). While this method could potentially lead to overestimating or underestimating the TRR, it allows us to observe general trends.

\begin{table}[ht]
\centering
\caption{Performance on Llama-2-7b with varying prompt template}
\label{tab:result_templates}
\begin{threeparttable}
\begin{tabular}{c|c|c|c}
\cline{1-4}
Prompt Template & TRR (Ours) & TRR (GCG) & TRR (ARCA) \\
\cline{1-4}   
Zero-shot & \textbf{0.98} & 0.96 & 0.94 \\
Alpaca & \textbf{0.98} & 0.90 & 0.64 \\
Fine-tuning \tnote{1} & 0.96 & 0.96 & 0.70 \\
ChatGPT & 0.92 & 0.92 & 0.54 \\
Tigerbot & \textbf{0.98} & 0.92 & 0.70 \\
Dolly V2 & \textbf{0.92} & 0.90 & 0.66 \\
\cline{1-4}
\end{tabular}
\begin{tablenotes}
    \item[1] This is the prompt template we customized for fine-tuning. See Section \ref{sec:setup}-4.
  \end{tablenotes}
  \end{threeparttable}
\end{table}

\begin{table*}[ht]
\centering
\caption{Performance on Llama-2-7b}
\label{tab:result_llama7b}
\begin{threeparttable}
\begin{tabular}{c|c|>{\centering\arraybackslash}p{2.5em}|>{\centering\arraybackslash}p{5em}|>{\centering\arraybackslash}p{2.5em}|>{\centering\arraybackslash}p{4em}|>{\centering\arraybackslash}p{4em}|>{\centering\arraybackslash}p{5em}}
    \cline{1-8}
    Suspect Model & Ground Truth \tnote{1} & TRR (Ours) & Difference \tnote{2} (Ours) & TRR (GCG) & Difference (GCG) & TRR (ARCA) & Difference (ARCA)\\ 
    \cline{1-8}
    \rowcolor[HTML]{ffe5e5} Llama-2-7b-chat & Positive & 0.18 & \textcolor{Red}{$\;\; 0.14 \uparrow$} & \textbf{0.22} & \textcolor{Green}{$\;\;0.06\;\;\;$} & 0.16 & \textcolor{Green}{$\;\;0.06\;\;\;$}\\
    \rowcolor[HTML]{ffe5e5} Vicuna-7b-v1.5 & Positive & 0.58  & \textcolor{Red}{$\;\; 0.54 \uparrow$} & \textbf{0.68} & \textcolor{Red}{$\;\;0.52 \uparrow$} & 0.52 & \textcolor{Red}{$\;\; 0.42 \uparrow$}\\
    \rowcolor[HTML]{ffe5e5} ELYZA-japanese-Llama-2-7b-instruct& Positive & \textbf{0.42} & \textcolor{Red}{$\;\; 0.38 \uparrow$} & 0.28 & \textcolor{Green}{$\;\;0.12\;\;\;$} & 0.22 & \textcolor{Red}{$\;\; 0.12 \uparrow$}\\ 
    \rowcolor[HTML]{ffe5e5} Llama2-Chinese-7b-Chat & Positive & 0.36 & \textcolor{Red}{$\;\; 0.32 \uparrow$} & \textbf{0.40} & \textcolor{Red}{$\;\;0.24 \uparrow$} & 0.28 & \textcolor{Red}{$\;\;0.18 \uparrow$}\\ 
    \rowcolor[HTML]{ffe5e5} Llama-2-7b-ft-instruct-es & Positive & \textbf{0.42} & \textcolor{Red}{$\;\; 0.38 \uparrow$} & 0.40 & \textcolor{Red}{$\;\;0.24 \uparrow$} & 0.34 & \textcolor{Red}{$\;\; 0.24 \uparrow$}\\
    \rowcolor[HTML]{ffe5e5} Meditron-7B & Positive & \textbf{0.46} & \textcolor{Red}{$\;\; 0.42 \uparrow$} & 0.42 & \textcolor{Red}{$\;\;0.26 \uparrow$} & 0.26 & \textcolor{Red}{$\;\;0.16 \uparrow$}\\
    \rowcolor[HTML]{ffe5e5} Orca-2-7b \tnote{3}  & Positive & \textbf{0.36} & \textcolor{Red}{$\;\; 0.32 \uparrow$} & 0.20 & \textcolor{Green}{$\;\;0.04\;\;\;$} & 0.20 & \textcolor{Green}{$\;\;0.10\;\;\;$}\\
    \rowcolor[HTML]{ffe5e5} Asclepius-Llama2-7B & Positive & 0.38 & \textcolor{Red}{$\;\; 0.34 \uparrow$} & \textbf{0.42} & \textcolor{Red}{$\;\;0.26 \uparrow$} & 0.28 & \textcolor{Red}{$\;\;0.18 \uparrow$}\\
    \rowcolor[HTML]{ecffe5} Mistral-7B-Instruct-v0.1 & Negative & \textbf{0.04} & \textcolor{Green}{$\;\; 0.00 \;\;\;$} & 0.06 & \textcolor{Green}{$-0.10\;\;\;\;$} & 0.08 & \textcolor{Green}{$-0.02\;\;\;\;$}\\
    \rowcolor[HTML]{ecffe5} Yi-6B-Chat & Negative & 0.00 & \textcolor{Green}{$-0.04 \;\;\;\;$} & 0.06 & \textcolor{Green}{$-0.10\;\;\;\;$} & 0.00 & \textcolor{Green}{$-0.10\;\;\;\;$}\\
    \rowcolor[HTML]{ecffe5} ChatGLM3-6B & Negative & \textbf{0.00} & \textcolor{Green}{$-0.04 \;\;\;\;$} & 0.04 & \textcolor{Green}{$-0.12\;\;\;\;$} & 0.06 & \textcolor{Green}{$-0.04\;\;\;\;$}\\
    \rowcolor[HTML]{ecffe5} Gemma-7b-it& Negative & 0.04 & \textcolor{Green}{$\;\; 0.00 \;\;\;$} & 0.04 & \textcolor{Green}{$-0.12\;\;\;\;$} & 0.04 & \textcolor{Green}{$-0.06\;\;\;\;$}\\
    \rowcolor[HTML]{ecffe5} Phi-2& Negative & \textbf{0.02} & \textcolor{Green}{$-0.02 \;\;\;\;$} & 0.16 & \textcolor{Green}{$\;\;0.02$} \tnote{4} $\;$ & 0.10 & \textcolor{Green}{$\;\;0.02\;\;\;$}\\
    \rowcolor[HTML]{ecffe5} OLMo-7B-Instruc & Negative & 0.04 & \textcolor{Green}{$\;\; 0.00 \;\;\;$} & 0.04 & \textcolor{Green}{$-0.12\;\;\;\;$} & 0.08 & \textcolor{Green}{$-0.02\;\;\;\;$}\\
    \rowcolor[HTML]{ecffe5} Falcon-7B-Instruct & Negative & \textbf{0.04} & \textcolor{Green}{$\;\; 0.00 \;\;\;$} & 0.14 & \textcolor{Green}{$-0.02$} \tnote{4} $\;\;$ & 0.08 & \textcolor{Green}{$-0.02\;\;\;\;$}\\
    \rowcolor[HTML]{ecffe5} GPT 3.5 & Negative & \textbf{0.02} & \textcolor{Green}{$-0.02 \;\;\;\;$} & 0.06 & \textcolor{Green}{$-0.10\;\;\;\;$} & 0.04 & \textcolor{Green}{$-0.06\;\;\;\;$}\\  
    \rowcolor[HTML]{ffffe5} CodeLlama-7b-Instruct \tnote{5} & Related & 0.06 & \textcolor{Green}{$\;\; 0.02 \;\;\;$} & 0.06 & \textcolor{Green}{$-0.10\;\;\;\;$} & 0.02 & \textcolor{Green}{$-0.08\;\;\;\;$} \\
    \rowcolor[HTML]{ffffe5} Llama-2-13b & Related & 0.10  & \textcolor{Red}{$\;\; 0.06 \uparrow$} & 0.34 & \textcolor{Red}{$\;\;0.18 \uparrow$} & 0.18 & \textcolor{Green}{$\;\;0.08\;\;\;$} \\
    \cline{1-8}
\end{tabular}
  \begin{tablenotes}
    \item[1] ``Positive'' indicates that the suspect model was fine-tuned from the original model. ``Negative'' indicates that the suspect model is unrelated to the original model. ``Related'' indicates that the suspect model was not fine-tuned from the original model but was trained by the owner of the original model under similar settings or datasets.
    \item[2] The difference between the TRR of the model and the highest TRR among other unrelated models. Values exceeding the \textbf{double} of the highest TRR among other unrelated models are highlighted in \textcolor{Red}{red $\uparrow$}, otherwise highlighted in \textcolor{Green}{green}.
    \item[3] The queries for evaluations are modified. See Section \ref{sec:setup}-3.
    \item[4] Two values would be \textcolor{Red}{$0.10 \uparrow$} and \textcolor{Red}{$0.08 \uparrow$} in case another model is not involved in the evaluation.
    \item[5] Though CodeLlama is reputedly based on Llama-2, it was trained in a cascaded manner for more than 500B tokens\cite{Roziere24:arXiv:codellama}. As training GPT-3 from scratch used around 300B tokens\cite{Brown20:arXiv:gpt3}, we consider CodeLlama as the same-family model instead of the fine-tuned model.
  \end{tablenotes}
\end{threeparttable}
\end{table*}

\begin{table}[ht]
\centering
\caption{Performance on Mistral-7B-v0.1}
\label{tab:result_mistral}
\begin{threeparttable}
\begin{tabular}{c|c|c|c}
\cline{1-4}
Suspect Model & Ground Truth & TRR & Difference \\
\cline{1-4}
    \rowcolor[HTML]{ffe5e5} Mistral-7B-Instruct-v0.1 & Positive & 0.14 & \textcolor{Red}{$\;\; 0.10 \uparrow$}\\
    \rowcolor[HTML]{ffe5e5} OpenHermes-2.5-Mistral-7B  & Positive & 0.32 & \textcolor{Red}{$\;\; 0.28 \uparrow$}\\
    \rowcolor[HTML]{ffe5e5} Dolphin-2.2.1-mistral-7b & Positive & 0.20 & \textcolor{Red}{$\;\; 0.16 \uparrow$}\\ 
    \rowcolor[HTML]{ffe5e5} Code-Mistral-7B& Positive & 0.28 & \textcolor{Red}{$\;\; 0.24 \uparrow$}\\ 
    \rowcolor[HTML]{ffe5e5} Hyperion-2.0-Mistral-7B  & Positive & 0.28 & \textcolor{Red}{$\;\; 0.24 \uparrow$}\\ 
    \rowcolor[HTML]{ffe5e5} Hermes-2-Pro-Mistral-7B & Positive & 0.22 & \textcolor{Red}{$\;\; 0.18 \uparrow$}\\
    \rowcolor[HTML]{ffe5e5} Mistral-7B-OpenOrca \tnote{1} & Positive & 0.28 & \textcolor{Red}{$\;\; 0.24 \uparrow$}\\
    \rowcolor[HTML]{ffe5e5} Starling-LM-7B-alpha & Positive \tnote{2} & 0.20 & \textcolor{Red}{$\;\; 0.16 \uparrow$}\\
    \rowcolor[HTML]{ecffe5} Llama-2-7b-chat & Negative & 0.02 & \textcolor{Green}{$-0.02 \;\;\;\;$}  \\
    \rowcolor[HTML]{ecffe5} Yi-6B-Chat & Negative & 0.02  & \textcolor{Green}{$-0.02 \;\;\;\;$} \\
    \rowcolor[HTML]{ecffe5} ChatGLM3-6B & Negative & 0.00 & \textcolor{Green}{$-0.04 \;\;\;\;$}\\
    \rowcolor[HTML]{ecffe5} Gemma-7b-it & Negative & 0.02 & \textcolor{Green}{$-0.02 \;\;\;\;$}  \\
    \rowcolor[HTML]{ecffe5} Phi-2 & Negative & 0.00 & \textcolor{Green}{$-0.04 \;\;\;\;$}\\
    \rowcolor[HTML]{ecffe5} OLMo-7B-Instruct & Negative & 0.02 & \textcolor{Green}{$-0.02 \;\;\;\;$}  \\
    \rowcolor[HTML]{ecffe5} Falcon-7B-Instruct & Negative & 0.04 & \textcolor{Green}{$\;\;0.02 \;\;\;$}\\
    \rowcolor[HTML]{ecffe5} GPT 3.5 & Negative & 0.02 & \textcolor{Green}{$-0.02 \;\;\;\;$}\\
    \rowcolor[HTML]{ffffe5} Mistral-7B-v0.2 & Related & 0.70 & \textcolor{Red}{$\;\; 0.66 \uparrow$} \\
    \cline{1-4}
    \end{tabular}
    \begin{tablenotes}
    \item[1] The queries for evaluations are modified. See Section \ref{sec:setup}.3.
    \item[2] Starling-LM-7B-alpha is fine-tuned based on Openchat 3.5, which is fine-tuned based on Mistral-7B-v0.1.
  \end{tablenotes}
  \end{threeparttable}
\end{table}

\subsection{Performance When the Prompt Template Changes} 
To investigate how decision boundaries change with variations in prompt templates, we assess Llama-2-7b using queries generated by our method, GCG, and ARCA on different prompt templates. The results are shown in Table \ref{tab:result_templates}. We found that the TRRs for both our queries and GCG's queries are not significantly affected when the prompt template changes, as these queries were generated using two templates simultaneously. Additionally, despite having more constraints than GCG, our method achieved better performance. Although ARCA achieved good performance on the zero-shot template, which is similar to the template used to generate its queries, its performance decreased dramatically when the prompt template changed. This indicates a shifting of the decision boundary with respect to the prompt template. Therefore, it is critical to use queries that are generalized on at least two prompt templates as fingerprints.

\subsection{Results on Existing Derived Models} 
\label{sec:results} 
We tested models derived from Llama-2-7b with queries generated by our method, GCG, and ARCA, and recorded the TRRs. We also tested models derived from Mistral-7B-v0.1 with queries generated by our method. The results are presented in Table \ref{tab:result_llama7b} and Table \ref{tab:result_mistral}, respectively. Both our method and GCG outperformed ARCA on fine-tuned models, demonstrating strong generalization abilities. While the TRRs of GCG are comparable to those of our method on fine-tuned models, our method significantly decreased the TRRs among unrelated models, thereby enabling queries to be used as fingerprints. Although all methods achieved an AUC score of 1.0 (excluding same-family models), because setting a definitive threshold for determination is impractical, our method showed a superior ability to effectively differentiate between derived and unrelated models, as the lowest TRR among derived models is more than three times the highest TRR among unrelated models. In contrast, GCG exhibited potential for false positives, and both GCG and ARCA exhibited potential for false negatives. 

Overall, we found that the TRRs reported by our method for most fine-tuned models are significantly higher than those for unrelated models, providing a clear indication of fine-tuning. Two official chat models, Llama-2-7b-chat and Mistral-7B-Instruct-v0.1, show relatively low TRRs among the fine-tuned models. Since the publishers of the original models have computational resources that other entities cannot afford, they may conduct much more intensive fine-tuning, thus leading to low TRRs.

Additionally, we found that the TRR for Llama-2-13b is relatively high, and the TRR for Mistral-7B-v0.2 is exceptionally high, despite they are not fine-tuned on original models. This is likely due to the similarity in training processes, structures, or the datasets used for model variations. Since only the owner of the original model is capable of training a similar model, this special case does not affect the applicability of our method.

\begin{figure}[ht]
  \centering
  \includegraphics[width=\linewidth]{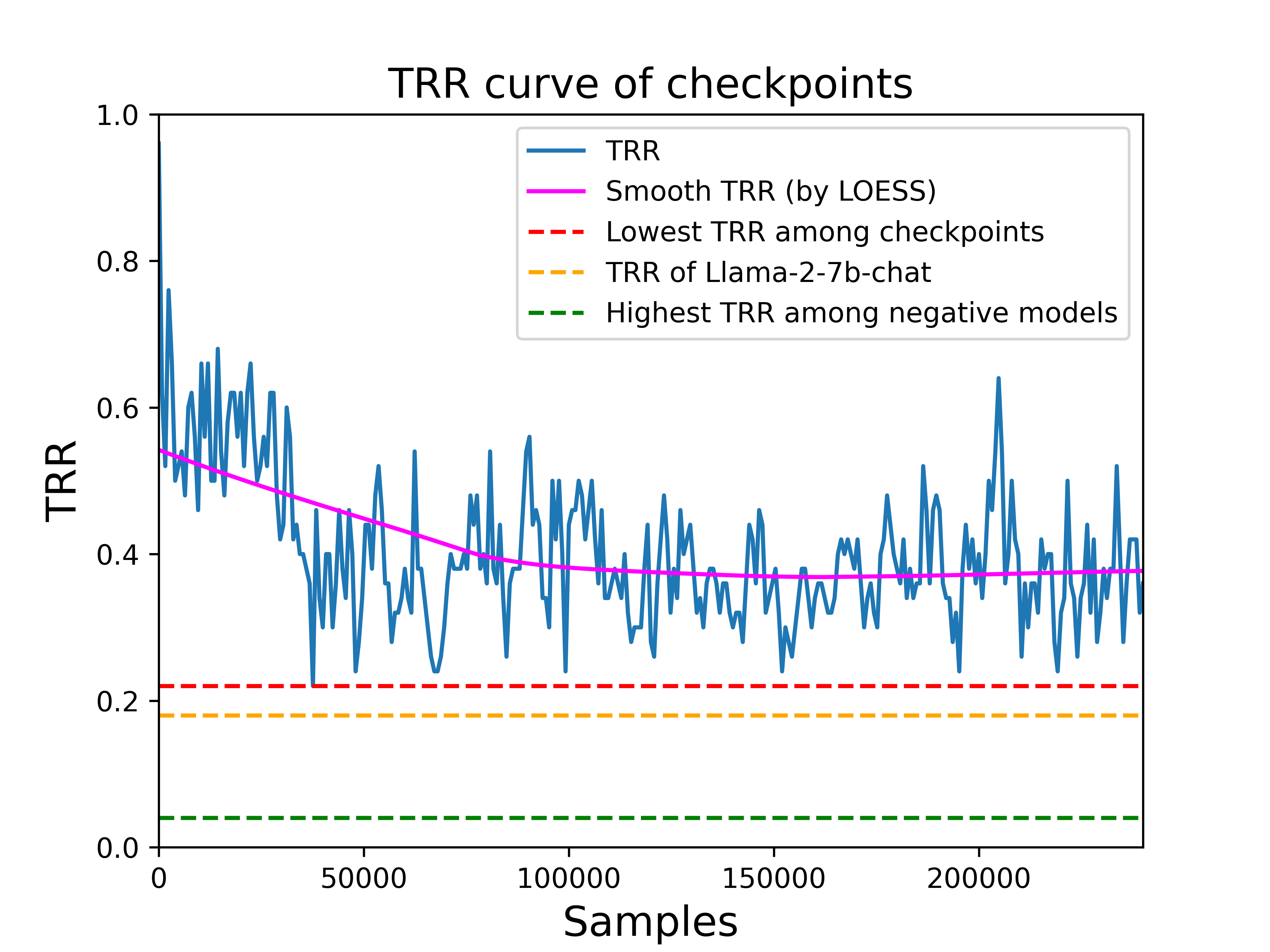}
  \caption{The TRR curve for all 240,000 samples of fine-tuning (\textcolor{Blue}{blue line}). We smooth it using LOESS (\textcolor{Magenta}{magenta line}), and compare it with the lowest TRR achieved (\textcolor{Red}{red line}), the TRR of Llama-2-7b-chat (\textcolor{Orange}{orange line}), and the highest TRR among unrelated models (\textcolor{Green}{green line}).}
  \label{fig:asr_all}
\end{figure}

\subsection{Performance on the Model We Fine-tuned}
We assessed all checkpoints of the model we fine-tuned using queries generated from the original Llama-2-7b and recorded TRRs. The TRR curve for all samples is shown in Fig. \ref{fig:asr_all}. We found that TRR dramatically decreased after fine-tuning on a few samples. For example, TRR decreased from 0.96 to 0.62 after fine-tuning on only 800 samples. However, after the first 40,000 samples, the TRR decrease slowed. These results suggest that to substantially reduce the effectiveness of \sysname, an attacker would need to engage in much more extensive fine-tuning, requiring considerable computational resources. 

Furthermore, we achieved the lowest TRR of 0.24 among all models fine-tuned from Llama-2-7b, with the exception of the officially fine-tuned Llama-2-7b-chat, since we fine-tuned the model with radical hyperparameters to maximize model change. Typically, for efficiency and performance, only a few targeted modules (e.g., only q\_proj and k\_proj), lower learning rates, or a linear learning rate scheduler are preferred.

\section{CONCLUSION}
In this work, we present \sysname, the first black-box fingerprinting-based copyright protection scheme for large language models. Unlike prior watermarking-based approaches, \sysname does not tamper the training process nor require additional processing, which enables \sysname to be employed in more complex situations and applied to models that have already been released. Our experiments on existing fine-tuned models validate the effectiveness of \sysname. Our fine-tuning results suggest that potentially bypassing \sysname's protection mechanisms would be exceedingly challenging.

\section*{Acknowledgments}
This work was supported in part by the Office of Naval Research under grant N00014-19-1-2621, and by the National Science Foundation under grants 2247560, 2154929, and 1916902.

\bibliographystyle{IEEEtran}
\bibliography{ref}



\end{document}